\documentclass[conference]{IEEEtran}
\usepackage{cite}

\ifCLASSINFOpdf
\usepackage[pdftex]{graphicx}
\else
\fi
\usepackage{amsmath}
\usepackage{url}

\begin{document}
%
\title{Extracting physical power plant parameters
from historical behaviour}

\author{\IEEEauthorblockN{David Kraljic, Miha Troha, Blaz Sobocan}\\
\IEEEauthorblockA{COMCOM d.o.o.\\
Idrija, Slovenia\\
Email: david.kraljic@comcom.si}


%


\thanks{
{\copyright} 2019 IEEE. Personal use of this material is permitted. Permission from IEEE must be obtained for all other uses, in any current or future media, including reprinting/republishing this material for advertising or promotional purposes, creating new collective works, for resale or redistribution to servers or lists, or reuse of any copyrighted component of this work in other works.}
}

\IEEEoverridecommandlockouts
\maketitle

\begin{abstract}
The information needed for fundamental modelling of the power markets -- the efficiency, start-up, fixed, and variable operating costs of each power plant -- is not publicly available. These parameters are usually estimated by considering the type of technology and the age of a power plant.

We present a method to extract these parameters for thermal power plants on the British electricity market using only the publicly available data. For each power plant, we solve a bilevel optimisation problem, where the inner level solves the Unit Commitment (UC) problem and outputs the optimal schedule given the prices of fuel, emissions, electricity, and the unknown plant parameters. The outer level then optimises over the plant parameters matching the historical production of each plant as closely as possible.
\end{abstract}


%
\IEEEpeerreviewmaketitle

\section*{Nomenclature}

\begin{IEEEdescription}[\IEEEusemathlabelsep\IEEEsetlabelwidth{$V_1,V_2,V_3$}]
\item[$t$] Time index
\item[$dt$] Time step between time indices [h]
\item[$P_t, P^{*}_t, \Tilde{P}_t$] Power [MW] at time $t$: continuous variable, optimal solution of the UC, observed value
\item[$\overline{P}_t$] Binary variable, one if power plant is on
\item[$\overline{C}_t$] Binary variable, one if power plant has started
\item[$m_t$] Maximum Export Limit (MEL) [MW]
\item[$r^{UP}, r^{DN}$] Ramping up and down limits [MW/h]
\item[$\eta$] Thermal efficiency
\item[$\sigma$] Start-up costs [\pounds]
\item[$\phi$] Fixed O\&M costs [\pounds /h]
\item[$\nu$] Variable O\&M costs [\pounds /MWh]
\item[$\epsilon$] Emission factor [tCO$_2$/MWh(fuel)]
\item[$w_t$] Wholesale electricity price [\pounds /MWh]
\item[$f_t$] Fuel price [\pounds /MWh(fuel)]
\item[$e_t$] Emissions price [\pounds /tCO$_2$]

\end{IEEEdescription}

\section{Introduction}

The fundamental models of power markets are widely used, for example, to predict electricity prices or to answer `what-if' questions such as the effect of phasing out of nuclear power or increase in renewables. These models are easily interpreted, capture the non-linear dependence of the prices on demand, correctly incorporate the effect of changing fuel and emission prices as well as intermittency of renewables. These properties are the reason for wide use of fundamental models in the power markets, despite the spread of ``machine learning'' models in other fields.

The key part of fundamental models is the supply curve (bid stack) - the supply of electricity ordered by the generators' cost of production. The price in a competitive market then results from the equilibrium of the demand and the supply and equals the marginal cost of the last power plant needed to satisfy demand. The renewables, whose production cannot be controlled, and nuclear plants, whose production does not change significantly, are usually subtracted from the demand, such that the supply curve only needs to model thermal units.

A good fundamental model, therefore, requires a detailed description of the thermal generators, which are usually modelled as a Unit Commitment problem \cite{UC}. The lack of information on exact physical and cost parameters of the generators, such as efficiency, start-up, fixed, and variable cost, results in simplified models. Generators are usually aggregated by fuel into a small number of units. Thermal efficiency is then estimated by considering the type of technology (e.g. simple cycle, combined cycle), fuel (e.g. coal, lignite, gas), and the age of power plants. Usually, only some of the cost types (fixed, variable, start-up) are considered and are assumed to be the same for plants of the same fuel type \cite{BRUNINX2013251, DEMODL, Schill2017}. 


We improve on the current approaches of determining plant parameters, by presenting a method to reverse engineer the physical parameters of power plants such that the historical production is closely matched, based only on the publicly available data. 

\section{Data}

We consider large thermal units on the UK market for the year 2018. These units have to submit production plans to the system operator at least one hour in advance of delivery. This data represents what power plants have committed to generate and thus we take these submitted volumes to represent power plants' position on the wholesale market. These large units also have to submit some of their physical and dynamic data to the system operator, such as ramping rate limits, stable export limits, and maximum export limits. This data is taken as an input to our UC model.

We use the following data sources:
\begin{itemize}
    \item Electricity prices: N2EX day-ahead auction \cite{N2EX}
    \item Gas prices: ICE UK daily gas future \cite{ICE}
    \item Coal prices: UK major producers self-reported price \cite{COAL}
    \item Emission prices: EEX + UK Carbon Price Floor \cite{EUAS1, EUAS2}
    \item Production: Production plans (FPN), ELEXON \cite{ELEXON}
    \item Export limits, ramping: Dynamic Data, ELEXON \cite{ELEXON}
    \item Emission factors \cite{emission}
\end{itemize}
 
\section{Problem description}
We are interested in obtaining the physical and cost parameters (the efficiency $\eta$, start-up costs $\sigma$, fixed operating costs $\phi$, and variable operating costs $\nu$) that minimise the error between the historical production $\Tilde{P}_{t}$ and the estimated production $P^{*}_{t}$. The estimated production is obtained by solving the UC problem for the given physical and cost parameters.

The described problem has a bilevel structure\cite{bilevelreview}:

\begin{align}
    &\text{outer level: }\quad \min_{\eta, \sigma, \phi, \nu} \quad \sum_{t} \left( P^{*}_t - \Tilde{P}_t \right)^2 \nonumber\\
    &\text{inner level: } \qquad \qquad \text{s. t. } P^{*}_t \in \arg \max \text{UC,} 
    \label{eqn:bilevel}
\end{align}

where the inner UC optimisation problem is:

\begin{align}
    \max_{P_t} \quad &\sum_t   P_t \left( w_t - \nu - f_t / \eta -  e_t \epsilon / \eta \right) dt - \overline{P}_t \phi \,dt - \overline{C}_t \sigma \nonumber \\
    \text{s. t.} &\qquad \overline{C}_{t+1} \geq \overline{P}_{t+1} - \overline{P}_{t} \quad \forall t \nonumber \\
    & \qquad -r^{DN} \leq P_{t+1} - P_{t} \leq r^{UP} \quad \forall t \nonumber \\
    & \qquad P_{t} \leq m_t \overline{P}_{t} \quad \forall t \nonumber \\
    & \qquad \dots
    \label{eqn:UC}
\end{align}
The first term in the UC objective represents the ``clean spark spread'', the second term the fixed costs, and the third term the start-up costs.
We only list the most relevant constraints. The actual formulation we use has several additional constraints (and variables) modelling the fact that thermal power plants have a ``Stable Export Limit'' below which they cannot produce at a constant level but must be in a ramp up or down phase.

\subsection{Assumptions}
Our modelling relies on a couple of assumptions regarding the operational and market behaviour of real power plants:
\begin{enumerate}
        \item Power plants run optimally by solving a UC problem of the form in Eq.\,\ref{eqn:UC}.
        
        UC models might not capture the physical properties or business logic of power plans completely. For example, power plants might operate based on in-house developed heuristics instead of solving an optimisation problem, they might run less frequently to postpone maintenance, they might provide reserve services instead of participating on the market, or they might take risks with respect to future or imbalance prices.

        \item Power plants buy fuel, emissions and sell electricity on the spot market (i.e. day-ahead or intraday). 
        
        Power plants can enter into long term contracts (e.g. seasonal, power purchasing agreements) for fuel, electricity, and emissions. Then they run such as to satisfy these obligations. The spot market assumption can be relaxed by including more market products for fuel, electricity etc. However, long term contracts are less transparent, therefore it is more difficult to estimate what prices the power plants get. 
\end{enumerate}

\subsection{Simplifications}
Taking the above assumptions as given, we also make certain simplifications in our UC modelling:

 \begin{enumerate}
    \item Efficiency is independent of part-load
    \item One ramping up (and down) rate across part-loads
    \item Optimise once for whole year on day-ahead prices (i.e perfect price foresight)
    \item One start-up cost (neglecting cold, warm, or more complicated start-up costs)
    \item No minimum and maximum up and down times
\end{enumerate}

The UC modelling assumptions can all be mitigated by including extra complexity, with additional decision variables and constraints. However, this comes at a cost. The underlying UC becomes more difficult to solve and our bilevel problem becomes exponentially more difficult. For example, adding cold start-up costs adds two dimensions to our search space on the outer level, namely the time after which cold start-up costs apply and the value of the cold start-up cost.

\section{Solving the bilevel problem}
The bilevel problem of Eqn.\,\eqref{eqn:bilevel} can be solved by multiple methods.
The simplest are direct search methods \cite{LEWIS2000191}, such as compass search, brute force, and Nelder-Mead.

The problem can also be solved with evolutionary or swarm algorithms, such as differential evolution, particle swarm, or simulated annealing \cite{swarm}.

The bilevel structure can also be transformed into a MILP \cite{Hart2017}, which is then solved with commercially available solvers with the solution transformed back into the original bilevel problem at the end. 

We have found that a combination of an evolutionary algorithm \cite{Storn1997} (to find an approximate location of minimum) followed by compass search works the fastest in practice as the computation can be massively parallelised. The inner UC problem is a standard MILP we solve using Gurobi\cite{gurobi}.

\subsection{The error landscape}
\begin{figure}[!ht]
    \centering
    \includegraphics[width=\columnwidth]{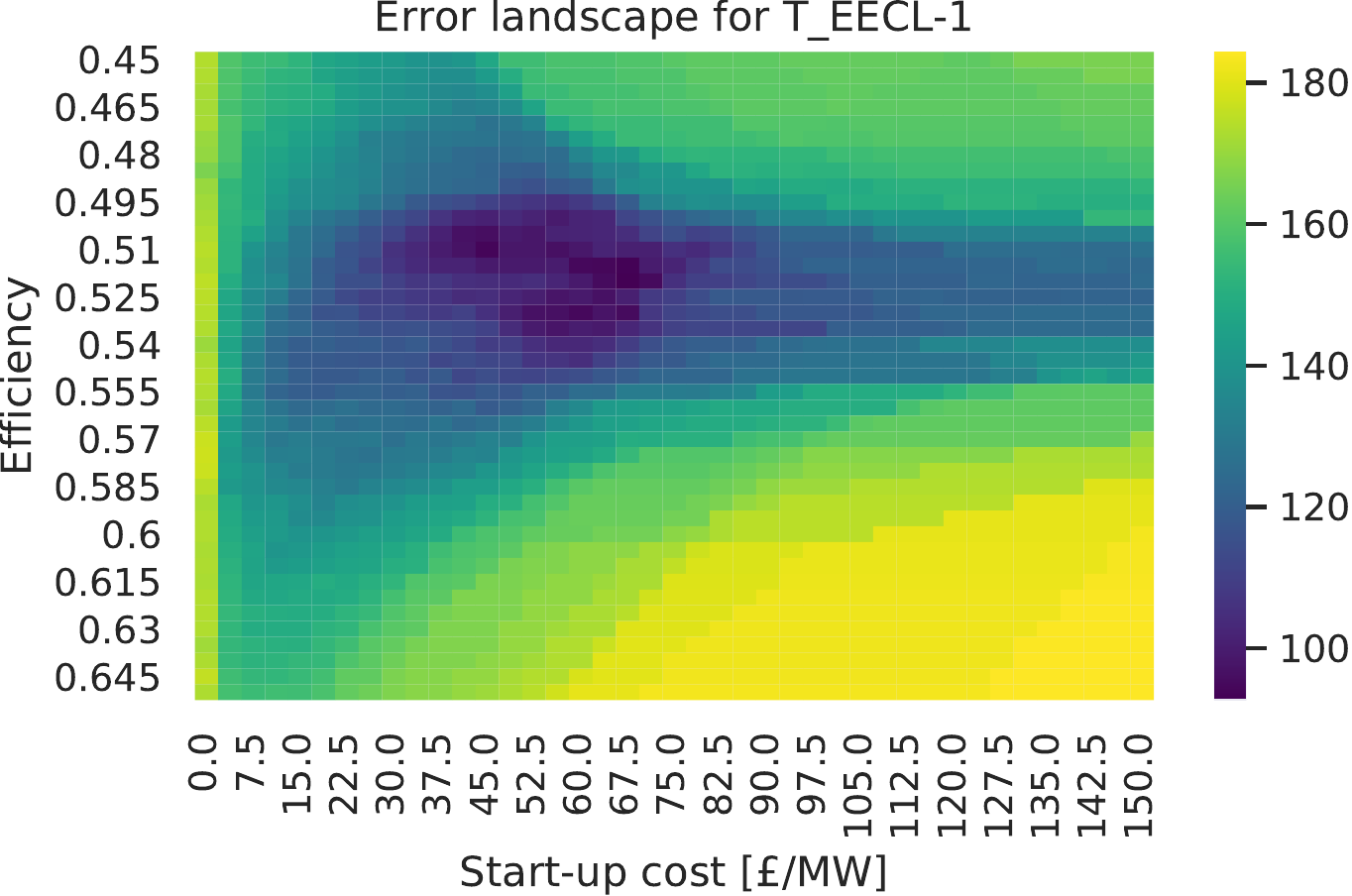}
    \caption{RMS Error [MW] for power plant T\_EECL-1 as a function of efficiency and start-up costs. In this figure we slice the parameter space by setting the fixed and variable costs to $\phi=11$\pounds/MW(cap) and $\nu=0$.}
    \label{fig:err}
\end{figure}

The objective of the outer level is a function of plant parameters ($\eta$, $\sigma$, $\phi$, $\nu$). The solution of our bilevel problem is a minimum of this function. We plot and illustrate this ``error landscape'' in Fig.\,\ref{fig:err}. As the search space is four-dimensional we can only display 2D slices through the landscape.

\section{Results}
\subsection{Efficiencies}
In this section we show the ability of our approach to discover best-fit power plant parameters consistent with average estimates appearing in literature, given the fuel and technology type as well as the age of power plants.

\begin{figure}[!ht]
    \centering
    \includegraphics[width=\columnwidth]{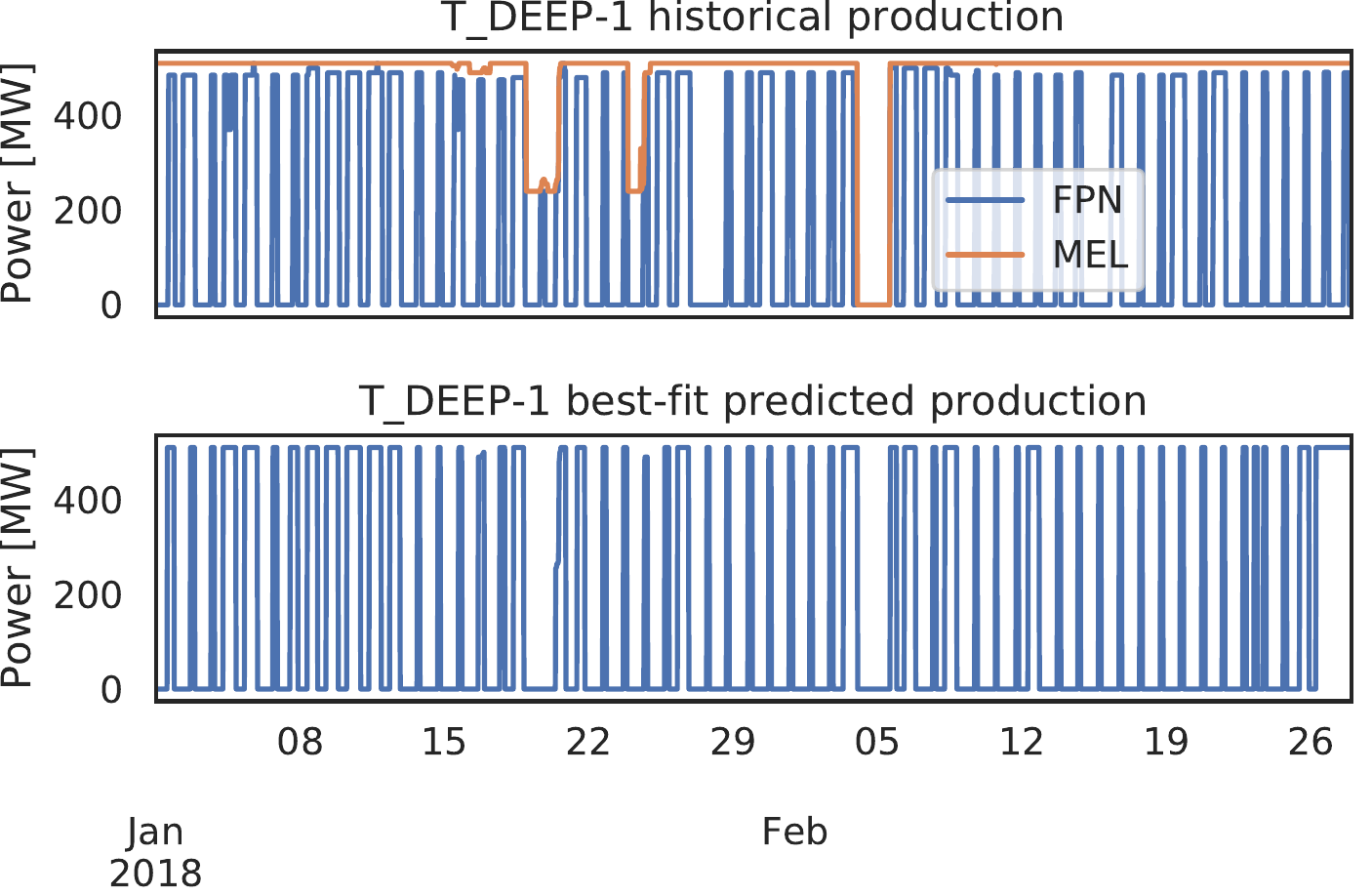}
    \caption{CCGT plant commissioned in 1994. Best-fit parameters are $\eta=0.53$, $\sigma=30$\pounds/MW(cap), $\phi=9.0$\pounds/h/MW(cap), $\nu=2.0$\pounds/MWh.}
    \label{fig:deep}
\end{figure}

\begin{figure}[!ht]
    \centering
    \includegraphics[width=\columnwidth]{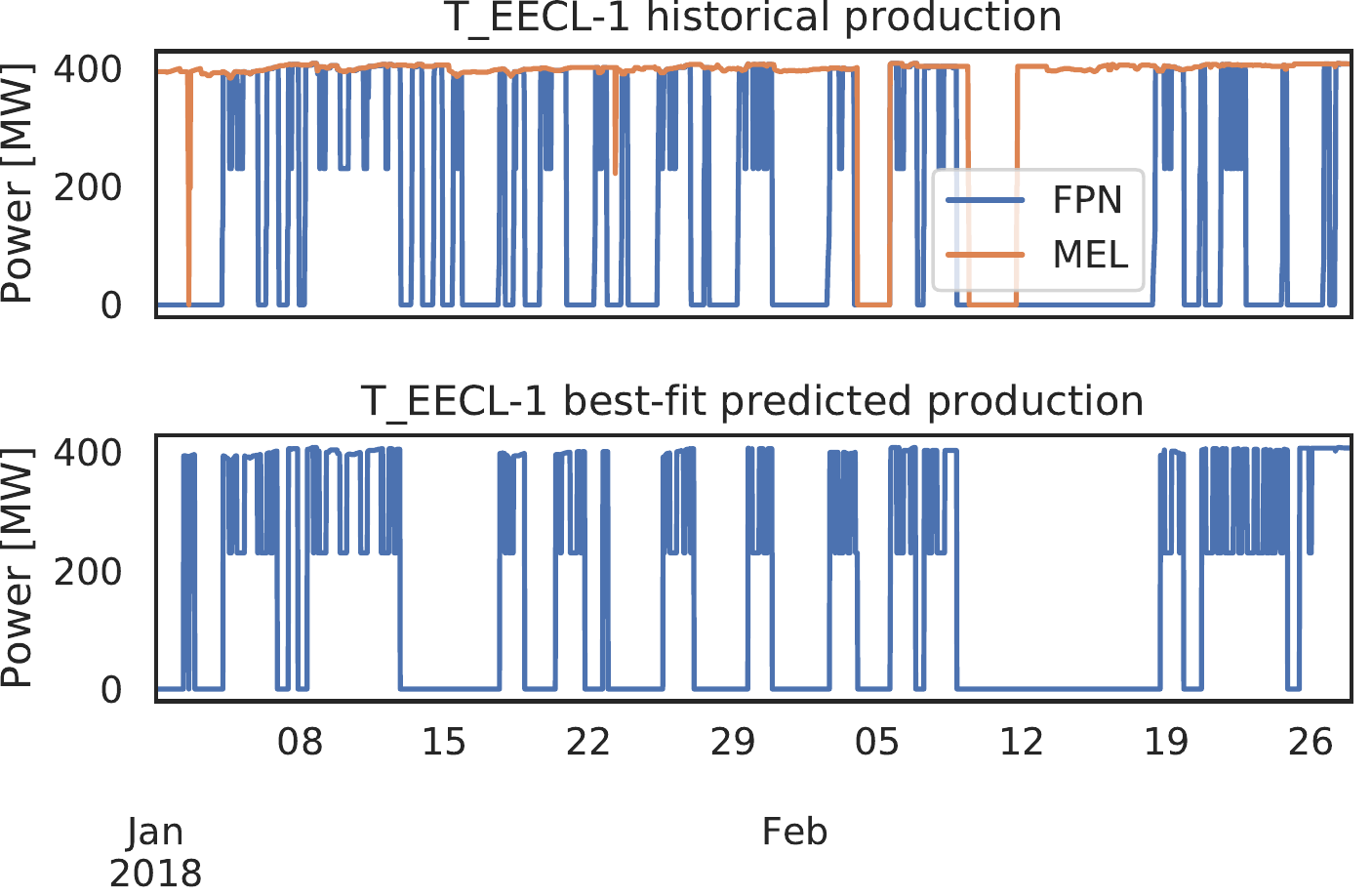}
    \caption{CCGT plant commissioned in 1999. Best-fit parameters are $\eta=0.53$, $\sigma=55$\pounds/MW(cap), $\phi=3.5$\pounds/h/MW(cap), $\nu=6.5$\pounds/MWh.}
    \label{fig:eecl}
\end{figure}

\begin{figure}[!ht]
    \centering
    \includegraphics[width=\columnwidth]{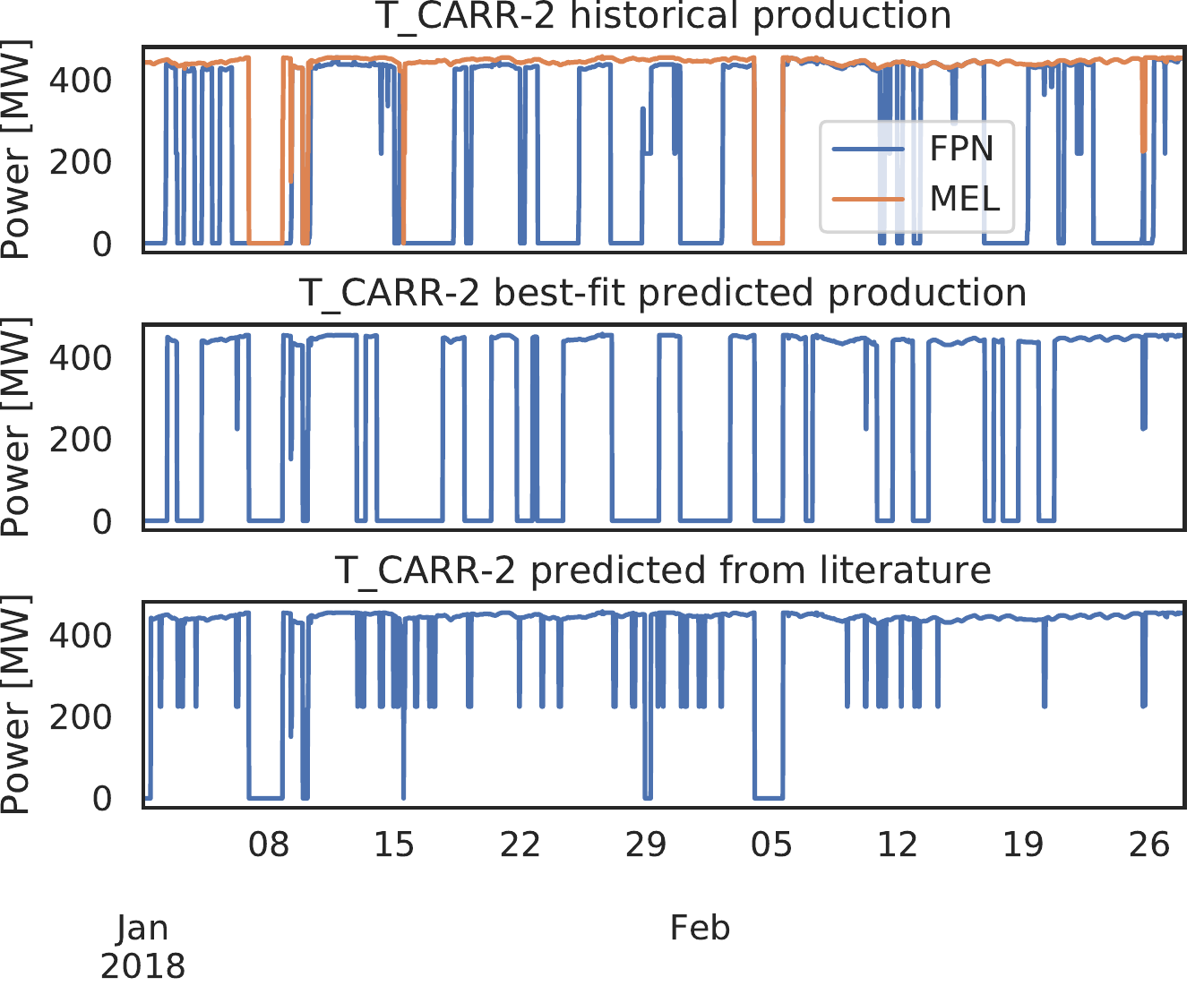}
    \caption{CCGT plant commissioned in 2016. Best-fit parameters are $\eta=0.58$, $\sigma=62$\pounds/MW(cap), $\phi=11.4$\pounds/h/MW(cap), $\nu=0.9$\pounds/MWh. Literature parameters from \cite{BRUNINX2013251, DEMODL, costs} are $\eta=0.57$, $\sigma=24$\pounds/MW(cap), $\phi=1.4$\pounds/h/MW(cap), $\nu=2.4$\pounds/MWh.}
    \label{fig:carr}
\end{figure}

Figs.\,\ref{fig:deep} and \ref{fig:eecl} are showing two CCGT power plants from late 1990s whereas Fig.\,\ref{fig:carr} shows a CCGT built in 2016. The best-fit efficiencies are about 0.53 for the older plants and 0.58 for the newer, with start-up costs of about 50\pounds/MW(cap), which is consistent with values for this type of technology and age group \cite{BRUNINX2013251, DEMODL, eff_cost_data_diw}.

\begin{figure}[!ht]
    \centering
    \includegraphics[width=\columnwidth]{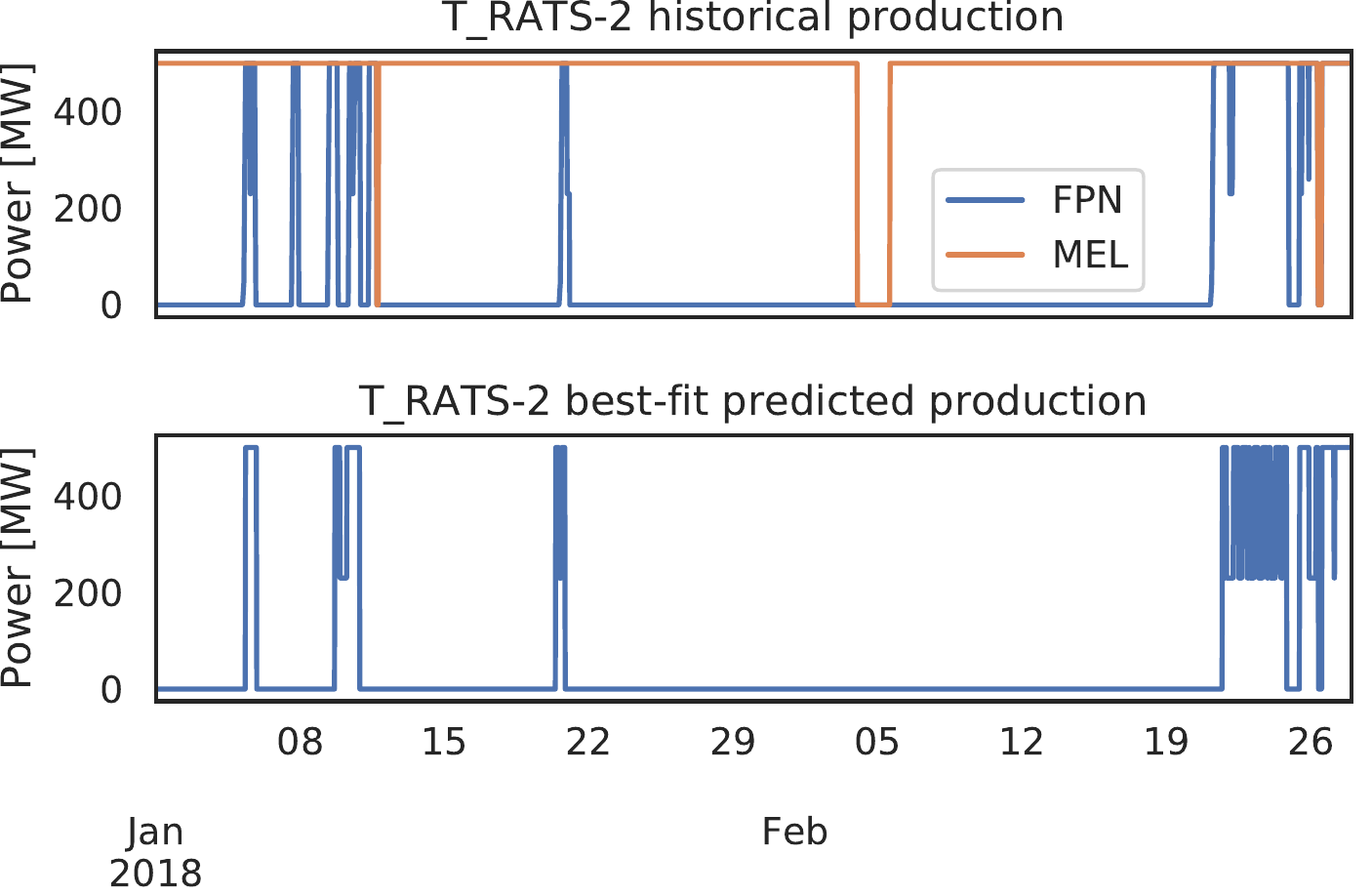}
    \caption{Coal plant commissioned in 1969. Best-fit parameters are $\eta=0.34$, $\sigma=76$\pounds/MW(cap), $\phi=4.8$\pounds/h/MW(cap), $\nu=0.0$\pounds/MWh.}
    \label{fig:rats}
\end{figure}

\begin{figure}[!ht]
    \centering
    \includegraphics[width=\columnwidth]{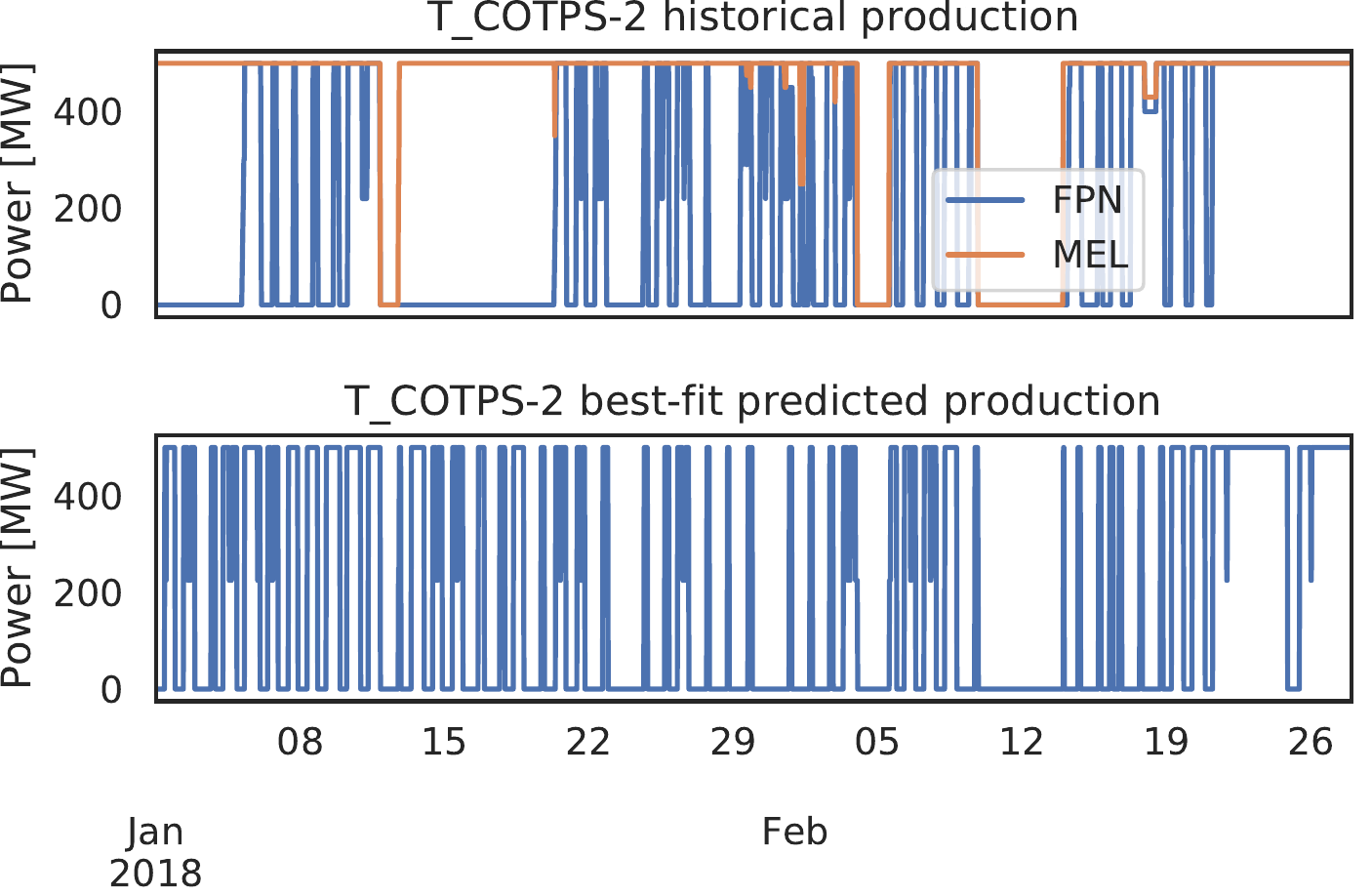}
    \caption{Coal plant commissioned in 1969, refurbished 2014. Best-fit parameters are $\eta=0.39$, $\sigma=30$\pounds/MW(cap), $\phi=5.0$\pounds/h/MW(cap), $\nu=2.0$\pounds/MWh.}
    \label{fig:cotps}
\end{figure}

Similarly, in Figs.\,\ref{fig:rats} and \ref{fig:cotps} we present two coal power plants. Both were commissioned in the late 1960s. However, only T\_RATS-1 has efficiency ($\eta=0.34$) consistent with the technology type and age. Power plant T\_COTPS-2 has undergone a refurbishment aimed at increasing efficiency\cite{COTPS_retrofit}, which might explain why the best-fit efficiency ($\eta=0.39$) is higher than expected for this plant.

\subsection{Costs}
We also show that power plants with the same efficiencies behave very differently, depending on the structure of their costs. For example, compare the very different production profiles of the two CCGTs in Figs.\,\ref{fig:deep} and \ref{fig:eecl}. 

Furthermore, we stress the importance of reverse engineering \emph{specific} power plant parameters. We show on the example of T\_CARR-2 (Fig.\,\ref{fig:carr}) that using the parameters appearing in the literature for this type and age of power plant leads to a very different prediction of plant production to the observed.

Therefore, the usual simplification in fundamental modelling where power plants of the same type have the same cost parameters leads to inaccurate fundamental models, especially in the short term  where electricity prices are often set by marginal costs of individual power plants.

\section{Discussion \& Conclusion}

We have shown that our bilevel framework of reverse engineering power plant parameters is capable of describing production profiles of real power plants. The obtained efficiencies and start-up costs are consistent with the average estimates found in the literature. 

The inner model (UC) of our bilevel set-up can be replaced with any other model (e.g. a linear model, decision tree, neural network). Then our bilevel framework becomes a standard set-up of ``machine learning'', where prediction error is minimised given some parametric model. Due to this similarity of our set-up and standard machine learning, the inner UC problem can be considered as an \emph{effective} model of power plants, where the deduced parameters might not be the ones of the real power plants, but the ones that fit their behaviour best given the specific UC model and historical data.

The advantages of using a UC model, besides interpretability, is that very complex behaviour can be specified with only a handful parameters (compared to hundreds in typical machine learning models) and there is no risk of overfitting the model. However, solving the UC problem and minimising the outer level error is computationally difficult, especially for more complex models (e.g. including minimum up/down times, several start-up costs, non-linear efficiency). This should be contrasted with e.g. linear models that have a closed-form solution, and fitting is performed in one step.

The framework we presented can be straightforwardly extended. The UC model can be made more complex by including extra physical parameters describing the power plants. Instead of assuming perfect foresight on the market prices, we can run the inner UC model in sequence on shorter time frames, simulating real-time behaviour. However, all this comes at a large computational cost.





%

\bibliographystyle{IEEEtran}
\bibliography{references.bib}


\end{document}